\documentclass[12pt]{article}  

\usepackage{amsmath,latexsym}

 \usepackage{graphicx}
\begin{document}
\begin{center}
{\LARGE \bf 
Thermodynamic stability in an Einstein universe}
\\ 
\vspace{1cm}
{\large  E. S. Moreira Jr.$^{a}$ 
\footnote{E-mail: {\tt moreira@unifei.edu.br}}	
	and J. P. A. Paula$^{b\,\,\,}$\footnote{E-mail:
{\tt almeida.paulo@ufabc.edu.br}}}
\\
\vspace{0.3cm}
{$^{a}$\em Instituto de Matem\'{a}tica e Computa\c{c}\~{a}o,}
{\em Universidade Federal de Itajub\'{a},}   \\
{\em Itajub\'a, Minas Gerais 37500-903, Brazil}  \\
\vspace{0.1cm}
{$^{b}$\em Centro de Ci\^{e}ncias Naturais e Humanas,}
{\em Universidade Federal do ABC,}   \\
{\em Santo Andr\'{e}, S\~{a}o Paulo 09210-580, Brazil}

\vspace{0.3cm}
{\large December, 2025}
\end{center}
\vspace{0.5cm}


\abstract{We calculate the Feynman propagator at finite temperature 
in an Einstein universe
for a neutral massive scalar field arbitrarily coupled to the Ricci curvature. Then, the propagator is used to determine the mean square fluctuation, the internal energy, and pressure of a scalar blackbody radiation as functions of the curvature coupling parameter $\xi$. By  studying thermodynamics of massless scalar fields, we show that the only value of $\xi$ consistent with stable thermodynamic equilibrium at 
all temperatures and for all radii of the universe 
is $1/6$, i.e., corresponding to the conformal coupling. 
Moreover, if electromagnetic and neutrino radiations
are present at the regime 
of high temperatures and/or large radii, 
we show that at least one scalar field 
must also be  present to ensure thermodynamic stability.

\vspace{0.5cm}

\section{Introduction}
\label{introduction}
Six decades ago, Penzias and Wilson found that a nearly 3K electromagnetic blackbody radiation fills up the cosmos
\cite{pen65}. As is well known, such a discovery has enormous implications, since it is related with the very beginning of our universe. Besides electromagnetic radiation, which is described by a massless vector field, the universe may house other types of radiation, such as scalar and spinor fields. 
Thus, it is certainly worth investigating the relationship between their thermodynamics and the topology and geometry of the universe.

In this work, the Einstein universe (EU) will be considered as a 
snapshot of a moment in time of a closed dynamical universe
\cite{par69,for75}. Its geometry is a static solution of the Einstein equations with line element given by
\footnote{Throughout the text fundamental constants are set equal to unity.}
\begin{equation}
	ds^{2}=dt^{2}-a^{2}[d\chi^{2}+\sin^{2}\chi (d\theta^{2}+\sin^{2}\theta\, d\varphi^{2})],	
	\label{le}
\end{equation}
i.e., the geometry of $R\times S^{3}$, where $a$ is the radius of the $3$-sphere, $\chi$ and $\theta$ $\in [0,\pi]$, and $\varphi$ $\in [0,2\pi]$, as usual. Quasi-Cartesian coordinates can be considered by taking 
$dx=a\sin\chi d\theta$, $dy=a\sin\chi\sin\theta d\varphi$ and
$dz=a d\chi$.

Quantum mechanics in the EU
goes back to Schr\"{o}dinger who has considered the covariant version of the wave equation
\begin{equation}
\left(\nabla_{\mu}\,\nabla^{\mu}+M^{2}\right)\phi({\rm x})=0	
\label{wave1}
\end{equation}
on $S^{3}$, to show that the energy spectrum  of a relativistic particle of mass $M$ is given by
\begin{equation}
\epsilon_{n} = \sqrt{\frac{n(n+2)}{a^{2}}+ M^2 },
\hspace{1.0cm}
n=0,1,2,\cdots,
\hspace{1.0cm}
g_{n}=(n+1)^{2}
\label{spectrum1}
\end{equation}
where $g_{n}$ is the degeneracy of each energy level $\epsilon_{n}$ \cite{sch38}.

Quantum fields in the EU
has been first addressed by Streeruwitz \cite{str75}, Ford \cite{for75},
and Mamaev {\it et al} \cite{mam76}. In this context of quantum fields in curved space, eq. (\ref{wave1}) is generalized to include
non minimal coupling with the Ricci curvature $R$ \cite{dav82},
\begin{equation}
	\left(\nabla_{\mu}\,\nabla^{\mu}+M^{2}+\xi R\right)\phi({\rm x})=0,	
	\label{wave2}
\end{equation}
where $R=6/a^{2}$ (see appendix \ref{geometry}).
The vacuum energy  density of a conformally invariant scalar field $\phi$ (i.e.,  $\xi=1/6$ and $M=0$ \cite{dav82}) has shown to be positive, namely \cite{for75} 
\begin{equation}
	\rho_{{\tt vacuum}}=\frac{1}{480\pi^{2}a^{4}},
\label{cved}	
\end{equation}
which contrasts with the well known negative vacuum energy density
$-\pi/6 L^{2}$ in $R\times S^{1}$ ($L$ is the length of the circle $S^{1}$. See refs. \cite{for75,dav82,ozc06}). The vacuum average of the stress-energy-momentum tensor
is diagonal,  
$\left<T^\mu{}_\nu\right>_{{\tt vacuum}}=
{\rm diag}(\rho_{{\tt vacuum}},-p_{{\tt vacuum}},-p_{{\tt vacuum}},-p_{{\tt vacuum}})$, and traceless due to conformal invariance. Then it results that
\begin{equation}
	p_{{\tt vacuum}}=\frac{1}{3}\rho_{{\tt vacuum}}
	\label{es1}	
\end{equation}
which is a radiation-like equation of state. It should be remarked that the Casimir pressure in eq. (\ref{es1}) is positive and therefore, from the thermodynamic point of view, it expands the EU, whereas the the Casimir pressure in $R\times S^{1}$ shrinks the circle.

Quantum fields at finite-temperature in EU has been boldly studied by Dowker and collaborators \cite{dow77,alt78}. By using the philosophy of obtaining radiation's thermodynamics through the ensemble average 
$\left<T^\mu{}_\nu\right>$ of a quantum field at temperature $T$ \cite{bro69}, the authors in ref. \cite{dow77} have shown that, for a conformally invariant scalar field $\phi$, the energy density is given by
\begin{equation}
\rho=	\rho_{{\tt vacuum}}+\rho_{T},
	\label{ced}	
\end{equation}
where 	$\rho_{{\tt vacuum}}$ is that in eq. (\ref{cved}).
It happens that, as $Ta\rightarrow 0$, $\rho_{T}$ in eq. (\ref{ced})
vanishes exponentially. Whereas, when $a\rightarrow\infty$ for a non vanishing $T$, $\rho_{T}$ approaches
exponentially $\pi^{2}T^{4}/30$, i.e., the well known blackbody expression.
At this point we should remark that, as the curvature $R$ equals $6/a^{2}$, 
by taking $a\rightarrow \infty$ any local quantity in the EU should become that in Minkowski spacetime. Conformal invariance also leads to 
\begin{equation}
	p=\frac{1}{3}\rho,
	\label{es2}	
\end{equation}
at all temperatures, as has been shown in ref. \cite{dow77} [cf. eq. (\ref{es1})].

Electromagnetic and neutrino radiations at temperature $T$ in the EU
have been addressed in ref. \cite{alt78}, showing that eqs. (\ref{ced}) and (\ref{es2}) still hold. Now the corresponding $\rho_{{\tt vacuum}}$ are
$11/240\pi^{2}a^{4}$ and $17/960\pi^{2}a^{4}$ for photons and neutrinos, respectively [cf. eq. (\ref{cved})]. The asymptotic behaviors of $\rho_{T}$
as $a\rightarrow\infty$ are now $\pi^{2}T^{4}/15$ and $7\pi^{2}T^{4}/60$, 
for photons and neutrinos, respectively, again known blackbody expressions (see, e.g., ref. \cite{kap06}).

An important issue studied in ref. \cite{alt78} concerns the comparison of $\rho_{T}$ in eq. (\ref{ced}) obtained from $\left<T^\mu{}_\nu\right>$, 
with $\rho_{T}$ that follows from statistical mechanics
\footnote{
For general use of the partition function see, e.g., textbook \cite{hua87}.} \cite{bre02,eli03,bez11}, i.e.,
\begin{equation}
\rho_{T}=\frac{1}{V}\sum_{n=0}^{\infty}\frac{g_{n}\epsilon_{n}}{e^{\epsilon_{n}/T}\pm 1}.
	\label{sm1}
\end{equation} 
For instance, in the case of minimally coupled ($\xi=0$) scalar bosons, eq. (\ref{spectrum1}) is used in eq. (\ref{sm1}), with the minus sign in the denominator, of course, and with $V$ given in eq. (\ref{volume}). 
Indeed, an equivalence has been shown, but only for conformally invariant fields \cite{alt78}. This detail will be relevant in the present study of thermodynamic stability in the EU, as we shall see.

Another interesting topic that has been investigated in the early works on
quantum fields in the EU is self-consistency. By feeding Einstein's equations with
$\left<T^\mu{}_\nu\right>$ that takes into account eqs. (\ref{cved}) and (\ref{es1}), it results a universe of Planck size \cite{for75,dow77}. Whereas, by feeding them with $\left<T^\mu{}_\nu\right>$ that takes into account eqs. (\ref{ced}) and (\ref{es2}) one gets $T$ as a function of $a$ \cite{alt78}. This is closely related to investigations that use $\left<T^\mu{}_\nu\right>$, in the context of backreaction,  
to possibly cure the well known dynamic instability of the EU (see, e.g., ref. \cite{her06}). 

These are essentially the main accounts and the basics we need to address our problem.
Other works will be considered in due course.

Going through the literature on thermodynamics of a scalar field $\phi$ in the EU,
we see no restriction on the value of the curvature coupling parameter $\xi$ in eq. (\ref{wave2}). Minimal ($\xi=0$) and conformal ($\xi=1/6$) couplings are the most commonly considered due to their typical properties: commitment to the strong equivalence principle and conformal invariance, respectively. In contrast, recent works in locally flat backgrounds have shown that the criteria for thermodynamic stability (see ref. \cite{cal85}) do constrain the range of possible values of $\xi$ \cite{lor15,mor17,mor20}. Here we investigate this issue for a background with non vanishing local curvature, $R\times S^{3}$.
The paper's plan is outlined below.

We present in appendix \ref{geometry} some elements obtained from eq. (\ref{le})
which are used in appendix \ref{thermal-green}
to calculate the Feynman propagator $G_{{\cal F}}({\rm x},{\rm x}')$ at finite temperature $T$ of a neutral and massive scalar field $\phi$, for arbitrary real values of the curvature coupling parameter $\xi$.
In section \ref{averages}, $G_{{\cal F}}({\rm x},{\rm x}')$ is renormalized and recast conveniently.
 Following the philosophy of refs. \cite{dow77,bro69} to derive thermodynamics of quantum fields from their thermal Green functions, 
in sections \ref{phi2} and \ref{stress}
we  work out from the renormalized
$G_{{\cal F}}({\rm x},{\rm x}')$
(by means of point-splitting \cite{dav82}) the ensemble averages $\left<\phi^{2}\right>$ and $\left<T^\mu{}_\nu\right>$.
A neat feature worth mentioning is that the propagator and all ensemble averages
are expressed in terms of the
modified Bessel functions of the second kind $K_{1}({\rm z})$ 
and $K_{2}({\rm z})$ \cite{arf85}.
Thermal properties of the ensemble averages are carefully investigated,
at all temperatures $T$ and for all radii $a$ of the universe,
and then used in section \ref{stability} to obtain the range over which $\xi$ can run without violating thermodynamic stability of a massless $\phi$. 
As an application of our overall results,
we address
the thermodynamic stability
of a mixed atmosphere of scalar bosons, neutrinos
and photons when $Ta\rightarrow\infty$. Final remarks are addressed in section \ref{conclusion}, closing the paper.

\section{Renormalized Feynman propagator}
\label{averages}
As has been mentioned in the previous section, in appendix \ref{thermal-green} we derive the Feynman propagator
$G_{{\cal F}}({\rm x},{\rm x}')$ in the E.U. in terms of  its related thermal Green function $G_{E}({\rm x},{\rm x}')$ [see eqs. (\ref{feynman}) and (\ref{ge3})]. As $\tau:=it$, we can analytically continue back to real values of $t$ in eq. (\ref{ge3}), resulting
\begin{eqnarray}
	G_{E}({\rm x},{\rm x}')=\frac{\mu}{4\pi^2\sin\alpha}
	\sum_{m=-\infty}^{\infty}\sum_{n=-\infty}^{\infty}(\alpha-2\pi n)
	\frac{K_{1}\left(\mu\sqrt{(\alpha-2\pi n)^2\, a^2-(t-t'-im\beta)^2}\right)}{\sqrt{(\alpha-2\pi n)^2\, a^2-(t-t'-im\beta)^2}}
	\label{ge4}
\end{eqnarray}
We should recall that $\alpha$ and $\mu$ are given in eqs. (\ref{angle}) and (\ref{mu}), respectively, and that $\beta$ is the inverse temperature,
i.e., $\beta:=1/T$.

At this point, it is worth presenting the asymptotic behaviors of 
the modified Bessel functions of the second kind $K_{\nu}({\rm z})$.
The following regimes will be explored in the text
\begin{eqnarray}
	&&K_{1}({\rm z}\rightarrow 0)=\frac{1}{{\rm z}}+\frac{{\rm z}}{2}\ln \frac{\rm z}{2}+\cdots, 
	\hspace{1.0cm}
	K_{2}({\rm z}\rightarrow 0)=\frac{2}{{\rm z}^2}-\frac{1}{2}+\cdots
	\label{ab1}
	\\
	&&K_{\nu}(|{\rm z}|\rightarrow\infty) = \sqrt{\frac{\pi}{2 {\rm z}}} e^{-{\rm z}}\left[1 + \frac{4\nu^2 - 1}{8{\rm z}} + \cdots \right]
	\label{ab2}
\end{eqnarray}
where ${\rm z}={\rm x}+i{\rm y}$, with {\rm x} and {\rm y} real numbers
\cite{abr65}.

By recasting $G_{E}({\rm x},{\rm x}')$ in eq. (\ref{ge4}) as
\begin{equation}
	G_{E}({\rm x},{\rm x}')=
	\sum_{m=-\infty}^{\infty}
	\sum_{n=-\infty}^{\infty}
	f^{(m,n)}_{({\rm x},{\rm x}')}
	\label{feynman2}
\end{equation}
we shall break up eq. (\ref{feynman2}) into the following contributions \cite{bro69}
\begin{eqnarray}
	&&
	G_{0}({\rm x},{\rm x}'):=f^{(0,0)}_{({\rm x},{\rm x}')},
	\hspace{2.6cm}
	G_{{\tt vacuum}}({\rm x},{\rm x}'):=\sideset{}{'}\sum_{n=-\infty}^{\infty}
	f^{(0,n)}_{({\rm x},{\rm x}')}
	\label{feynman3}
	\\
	&&
	G_{{\tt thermal}}({\rm x},{\rm x}'):=\sideset{}{'}\sum_{m=-\infty}^{\infty}
	f^{(m,0)}_{({\rm x},{\rm x}')},
	\hspace{0.6cm}
	G_{{\tt mixed}}({\rm x},{\rm x}'):=\sideset{}{'}\sum_{m=-\infty}^{\infty}\,
	\sideset{}{'}\sum_{n=-\infty}^{\infty}
	f^{(m,n)}_{({\rm x},{\rm x}')}
	\label{feynman4}
\end{eqnarray}
where the prime indicates that the term corresponding to $m=0$ or $n=0$
in the summation should be discarded. Recalling that $\alpha$ equals
$s/a$, where $s$ is the geodesic distance, by setting $a\rightarrow\infty$
while keeping both $s$ and $\xi$ [see eq. (\ref{mu})] fixed, one sees from eq. (\ref{ge4}) that 
$G_{0}$ in eq. (\ref{feynman3}) approaches progressively the familiar Minkowski expression. Accordingly, following refs. \cite{for75} and \cite{dow77}, such a contribution will be dropped resulting that  
\begin{equation}
	G({\rm x},{\rm x}'):=
	G_{{\tt vacuum}}({\rm x},{\rm x}')+G_{{\tt mixed}}({\rm x},{\rm x}')+G_{{\tt thermal}}({\rm x},{\rm x}')
	\label{rfeynman}
\end{equation}
is the renormalized $G_{E}$
\footnote{A similar approach has been used in ref. \cite{mor20} in the context of $R\times S^{1}$.}.

Noting eq. (\ref{rfeynman}) and
taking into account eqs. (\ref{ab1}) and (\ref{ab2}), we see that 
$G_{{\tt vacuum}}$ does not depend on $T$
and vanishes if $a\rightarrow\infty$,
$G_{{\tt thermal}}$
vanishes if $T\rightarrow 0$, and that
$G_{{\tt mixed}}$
vanishes if $a\rightarrow\infty$ or if $T\rightarrow 0$ alike.
These asymptotic behaviors of the terms in eq. (\ref{rfeynman}) justify their subscripts.

\section{$\left<\phi^{2}\right>$}
\label{phi2}
The mean square fluctuation $\left<\phi^{2}\right>$ is an important physical quantity. It is a measure of how much $\phi$ fluctuates  around its mean value zero, and also appears as part of other quantities.
We will present some details of its calculation since it pedagogically illustrates how to obtain more complicated ensemble averages such as $\left<T^\mu{}_\nu\right>$. Moreover, when asymptotic regimes of temperature and size of the universe 
are considered, $\left<\phi^{2}\right>$ and $\left<T^\mu{}_\nu\right>$
share similar features, especially regarding to their dependence on $\xi$:
if $\xi<1/6$, $\xi=1/6$, or $\xi>1/6$.

Formally $\left<\phi^{2}\right>$ can be calculated by taking minus the imaginary part of $G_{{\cal F}}({\rm x},{\rm x}')$ as ${\rm x}'\rightarrow{\rm x}$ \cite{ful89}. In other words, by noticing eq. (\ref{feynman}) and the renormalized $G_{E}$ in eq. (\ref{rfeynman}),
it follows that 
\begin{eqnarray}
&&
\left<\phi^{2}\right>=\lim_{{\rm x}'\rightarrow{\rm x}}	
\Re\left(G({\rm x},{\rm x}')\right)
\label{defphi2}
\\
&&
\hspace{0.9cm}
=\left<\phi^{2}\right>_{{\tt vacuum}}+
\left<\phi^{2}\right>_{{\tt mixed}}+
\left<\phi^{2}\right>_{{\tt thermal}}
\label{termsphi2}	
\end{eqnarray}
Clearly, $\left<\phi^{2}\right>_{{\tt vacuum}}$ is independent of $T$ 
vanishing if $a\rightarrow\infty$,
$\left<\phi^{2}\right>_{{\tt mixed}}$
vanishes if $a\rightarrow\infty$ or  if $T\rightarrow 0$, and
$\left<\phi^{2}\right>_{{\tt thermal}}$
vanishes if $T\rightarrow 0$.

\subsection{$\left<\phi^{2}\right>_{{\tt vacuum}}$}
\label{vphi2}
Noting eq. (\ref{angle}) and taking 
$\theta=\theta'$ and
$\varphi=\varphi'$, it results that 
$\alpha=\Delta\chi:=\chi-\chi'$.
Further one makes $t=t'$ and expands $f^{(0,n)}_{({\rm x},{\rm x}')}$ [see eq. (\ref{feynman3})]
in powers of $\Delta\chi$. The procedure of building up such a series involves the derivative of $K_{1}({\rm z})$
which brings $K_{2}({\rm z})$ to the problem
\cite{gra07,mat17}.
Thus, according to eq. (\ref{defphi2}), one sets $\Delta\chi=0$, 
obtaining that the vacuum fluctuation is given by
\begin{eqnarray}
&&	 \left<\phi^{2}\right>_{{\tt vacuum}}= 
	\frac{1}{2\pi^2a^2}\Re\Bigg(\sum_{n = 1}^{\infty}\bigg[
	\frac{\sqrt{M^2 a^2+6 \xi -1}}{2 \pi  n}K_1\left(2\pi n\sqrt{M^2 a^2+6 \xi -1}\right)
	\hspace{2.0cm}
	\nonumber
	\\ 
	&&\hspace{4.0cm} -\left(M^2 a^2+6 \xi -1\right)K_2\left(2\pi n  \sqrt{M^2 a^2+6 \xi -1}\right)\bigg]\Bigg)
	\label{vphi2-1}
\end{eqnarray}
Now using eq. (\ref{ab1}) in eq. (\ref{vphi2-1}) for a conformally invariant $\phi$
and noticing the appearance of
$\zeta(2)=\pi^2/6$, one quickly sees that
\begin{equation}
\left<\phi^{2}\right>_{{\tt vacuum}}= -\frac{1}{48\pi^2 a^2},
\hspace{3.0cm} 
M=0, \hspace{0.2cm} \xi=\frac{1}{6}
\label{vphi2-2}
\end{equation}
in agreement with a recent work on quantum Brownian motion in the EU \cite{fer24}. As expected, $\left<\phi^{2}\right>_{{\tt vacuum}}$
in eq. (\ref{vphi2-2}) vanishes when $a\rightarrow\infty$. 
Although a detailed analyses on how vacuum fluctuations vary with $\xi$
is rather outside the scope of our study,
Figure \ref{fig1} illustrates how $\left<\phi^{2}\right>_{{\tt vacuum}}$ varies with $\xi$ around $1/6$
for a massless $\phi$ in an EU with unitary radius.
\begin{figure}[htbp]
	\centering
	\includegraphics[scale=0.6]{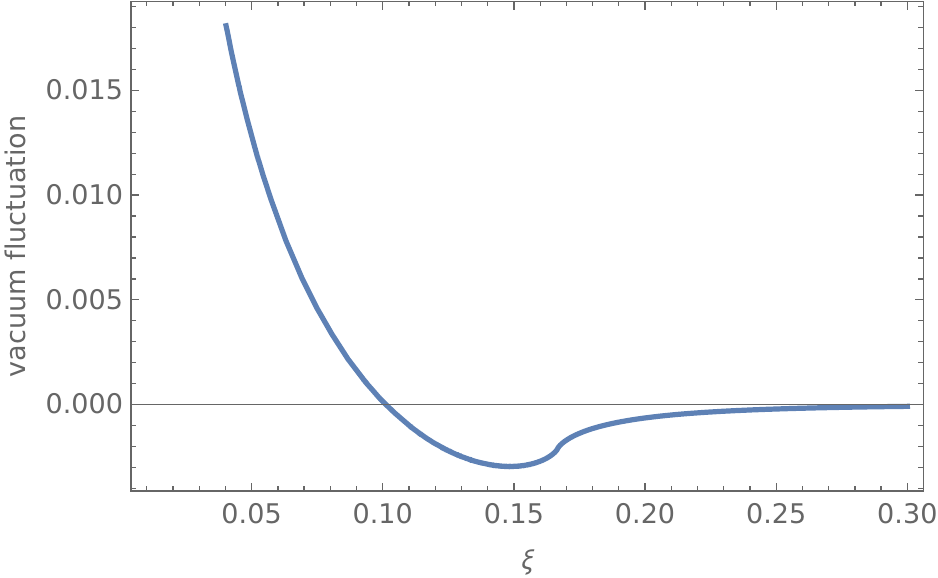}
	\caption{\footnotesize{$\left<\phi^{2}\right>_{{\tt vacuum}}$ vs. $\xi$, for $\xi>0$, $M=0$ and $a=1$. As $\xi\rightarrow\infty$ $\left<\phi^{2}\right>_{{\tt vacuum}}$ approaches zero from below. As $\xi\rightarrow 0^{+}$ it grows indefinitely.}} 
	\label{fig1}
\end{figure}

\subsection{$\left<\phi^{2}\right>_{{\tt thermal}}$}
\label{tphi2}
The procedure to calculate $\left<\phi^{2}\right>_{{\tt thermal}}$}
in eq. (\ref{termsphi2})
is identical to that for $\left<\phi^{2}\right>_{{\tt vacuum}}$ described above. Now the term carrying $K_{2}({\rm z})$ is eliminated when 
${\rm x}'\rightarrow{\rm x}$, yielding
\begin{eqnarray}
	&&	 \left<\phi^{2}\right>_{{\tt thermal}}= 
	\frac{T}{2\pi^2a}\Re\left(\sum_{m = 1}^{\infty}
	\left[
	\frac{\sqrt{M^2 a^2+6 \xi -1}}{m}
	K_1\left(\frac{m\sqrt{M^2 a^2+6 \xi -1}}{Ta}\right)
	\right]
	\right)
    \label{tphi2-1}
\end{eqnarray}
When $\xi=1/6$,  eq. (\ref{tphi2-1}) becomes the blackbody vacuum fluctuation in Minkowski spacetime and, of course, it does not depend on $a$. Indeed, noting eq. (\ref{ab1}), it follows the well know massless result
\begin{equation}
	\left<\phi^{2}\right>_{{\tt thermal}}= \frac{T^2}{12},
	\hspace{3.0cm} 
	M=0, \hspace{0.2cm} \xi=\frac{1}{6}
	\label{tphi2-2}
\end{equation}
which is also the leading contribution when $M$ and $\xi$ are arbitrary; but $Ta\rightarrow \infty$. 

\begin{figure}[htbp]
	\centering
	\includegraphics[scale=0.8]{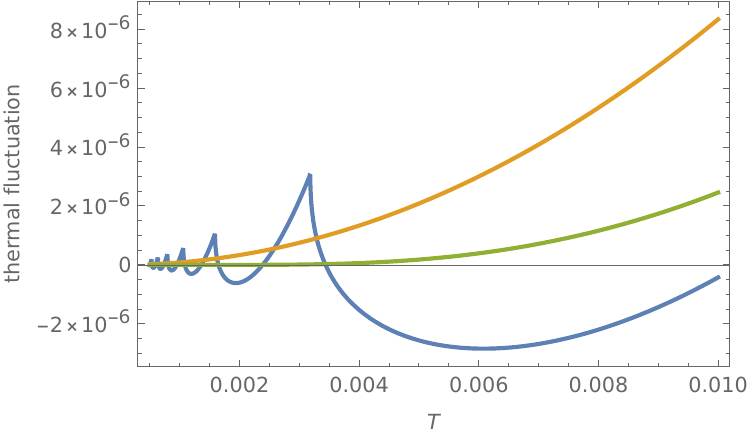}
	\caption{\footnotesize{$\left<\phi^{2}\right>_{{\tt thermal}}$ vs. $T$, for $M=0$ and $a=1$. On the right: the lower (blue) plot,
	the middle (green) plot, and the upper 	(yellow) plot correspond,
	respectively, to $\xi=0.1666$,
    $\xi=0.1667$, and $\xi=1/6$. Note that these asymptotic behaviors
    as $T\rightarrow 0$ differ radically from each other though their  associated values of $\xi$ are not that different.}}
	\label{fig2}
\end{figure}

It is vital to determine how $\left<\phi^{2}\right>_{{\tt thermal}}$
behaves for $\xi\neq 1/6$ as $Ta\rightarrow 0$, since it carries the hallmark that prevents stable thermodynamic equilibrium when $\xi<1/6$ as we shall see.
Figure \ref{fig2} shows the thermal fluctuation's behavior when $\xi<1/6$ (oscillatory cusp-like),
$\xi=1/6$ (quadratic), and when $\xi>1/6$ (exponential), at low temperatures for an EU of unitary radius. Although at this regime of low temperatures they contrast among themselves, they do join each other when  $T\rightarrow \infty$, corresponding to eq. (\ref{tphi2-2}).

In order to see how the plots in Figure \ref{fig2} for $\xi\neq 1/6$ arise we must consider eq. (\ref{ab2}) in eq. (\ref{tphi2-1}).The argument applies also to massive fields when $a$ is small enough; but we will take $M=0$ to simplify. When $Ta\rightarrow 0$, it is straightforward to show that
\begin{equation}
	\left<\phi^{2}\right>_{{\tt thermal}}= \left(\frac{T}{2\pi}\right)^{3/2}
    \left(\frac{\sqrt{1-6\xi}}{a}\right)^{1/2}\sum_{m=1}^{\infty}m^{-3/2}
    \cos\left(\frac{m\sqrt{1-6\xi}}{Ta}-\frac{\pi}{4}\right)+\cdots,
	\hspace{0.3cm} \xi<\frac{1}{6}
	\label{tphi2-3}
\end{equation}
and that  
\begin{equation}
	\left<\phi^{2}\right>_{{\tt thermal}}= \left(\frac{T}{2\pi}\right)^{3/2}
	\left(\frac{\sqrt{6\xi-1}}{a}\right)^{1/2}e^{-\sqrt{6\xi-1}/Ta}+\cdots,
	\hspace{0.5cm} \xi>\frac{1}{6}.
	\label{tphi2-4}
\end{equation}
Eq. (\ref{tphi2-3}) explains why $\left<\phi^{2}\right>_{{\tt thermal}}$
in Figure \ref{fig2} oscillates in a cusp-like manner, acquiring negative and positive values, before vanishing at the absolute zero of temperature. 
The contrast between eqs. (\ref{tphi2-3}) and (\ref{tphi2-4}) resides mainly in the fact that, when $\xi<1/6$, ${\rm z}$ in $K_{1}({\rm z})$ becomes imaginary.

\subsection{$\left<\phi^{2}\right>_{{\tt mixed}}$}
\label{mphi2}
We turn to the calculation of $\left<\phi^{2}\right>_{{\tt mixed}}$
in eq. (\ref{termsphi2}) by means of the now familiar procedure
that led to the other terms. Noticing eq. (\ref{feynman4}), we find that
\begin{eqnarray}
	&&	 \left<\phi^{2}\right>_{{\tt mixed}}= 
	\frac{T}{\pi^2a}\Re\Bigg(\sum_{m,n = 1}^{\infty}
	\Bigg[
	\frac{\sqrt{M^2 a^2+6 \xi -1}}{\sqrt{(2\pi n Ta)^{2}+m^{2}}}K_1\left(\frac{\sqrt{M^2 a^2+6 \xi -1}}{Ta}\sqrt{(2\pi n Ta)^{2}+m^{2}}\right)
	\nonumber
	\\ 
	&&\hspace{1.0cm} -\frac{\left(M^2 a^2+6 \xi -1\right)(2\pi n)^{2}Ta}{(2\pi n Ta)^{2}+m^{2}}K_2\left(\frac{ \sqrt{M^2 a^2+6 \xi -1}}{Ta}\sqrt{(2\pi n Ta)^{2}+m^{2}}\right)\Bigg]\Bigg).
	\label{mphi2-1}
\end{eqnarray}
Clearly, as happens to  $\left<\phi^{2}\right>_{{\tt vacuum}}$, the expression in
eq. (\ref{mphi2-1}) vanishes when $a\rightarrow\infty$. Also it vanishes
when $T\rightarrow 0$, as happens to $\left<\phi^{2}\right>_{{\tt thermal}}$.
We further notice that
\begin{eqnarray}
	&&	 \left<\phi^{2}\right>_{{\tt mixed}}= 
	\frac{T^2}{\pi^2}\sum_{m,n=1}^{\infty}
	\frac{m^2-(2\pi n Ta)^{2}}{[m^{2}+(2\pi n Ta)^{2}]^2},
	\hspace{2.0cm} M=0, \hspace{0.2cm} \xi=\frac{1}{6}
\label{mphi2-2}	
\end{eqnarray}
where once again eq. (\ref{ab1}) has been used.
In fact the double series in eq. (\ref{mphi2-2})
is not absolutely convergent. The order in which the summations are evaluated matters. Such an ambiguity  
is also present in calculations of $\left<T^\mu{}_\nu\right>$
in $R\times S^{1}$ \cite{mor20}
\footnote{Perhaps we should remark that it will not affect our study on thermodynamic stability in $R\times S^3$.}.

By summing first over $n$ in eq. (\ref{mphi2-2}) it results \cite{mat17}
\begin{equation}
	\left<\phi^{2}\right>_{{\tt mixed}}=-\frac{T^2}{12}+
	\frac{1}{8\pi^2 a^{2}}
	\sum_{k=1}^{\infty}{\rm cosech}^{2}\left(\frac{k}{2Ta}\right)
	\label{mphi2-3}
\end{equation}
whereas by summing first over $m$ one must add $T/4\pi^2 a$ to the RHS
of eq. (\ref{mphi2-3}). It is worth mentioning that a physical argument in favor of ``summing first over $n$'' is that, in $R\times S^{1}$,
it is consistent with the third law of thermodynamics (see ref. \cite{mor20} and references therein).
By manipulating the summation in eq. (\ref{mphi2-3}) as in ref. \cite{mor20}
two asymptotic behaviors arise, namely
\begin{eqnarray}
&&
\left<\phi^{2}\right>_{{\tt mixed}}=-\frac{T^2}{12}+
\frac{1}{2\pi^2 a^{2}}e^{-1/Ta}+\cdots, \hspace{2.5cm} Ta\ll 1
\label{mphi2-4}
\\
&&
\left<\phi^{2}\right>_{{\tt mixed}}=-\frac{T}{4\pi^2 a}+
\frac{1}{48\pi^2 a^{2}}
-2T^2 e^{-4\pi^2 Ta}+\cdots, \hspace{0.7cm} Ta\gg 1.
\label{mphi2-5}
\end{eqnarray}
We notice that the leading contribution in Eq. (\ref{mphi2-4}) is $-\left<\phi^{2}\right>_{{\tt thermal}}$
[cf. eq. (\ref{tphi2-2})], and that the subleading contribution 
in eq. (\ref{mphi2-5}) is $-\left<\phi^{2}\right>_{{\tt vacuum}}$
[cf. eq. (\ref{vphi2-2})].

If $\xi\neq 1/6$, but close enough to $1/6$, eqs. (\ref{mphi2-4}) and (\ref{mphi2-5}) still hold. Now, when one takes $Ta\rightarrow 0$, 
the behaviors of 
$\left<\phi^{2}\right>_{{\tt mixed}}$ with $T$
for $\xi<1/6$, $\xi=1/6$ [eq. (\ref{mphi2-4})] and $\xi>1/6$
differ radically again, in particular with that corresponding to $\xi<1/6$ presenting the oscillatory cusp-like pattern
of $\left<\phi^{2}\right>_{{\tt thermal}}$
in Figure \ref{fig2}, but now with the important difference that amplitudes are typically much larger
\footnote{It results that $\left<\phi^{2}\right>$ in eq. (\ref{termsphi2}) shows an oscillatory cusp-like pattern when $Ta\rightarrow 0$ and $\xi<1/6$.}. 
For $\xi>1/6$, $-\left<\phi^{2}\right>_{{\tt mixed}}$ exponentially decays as $Ta\rightarrow 0$ like $\left<\phi^{2}\right>_{{\tt thermal}}$.

\subsection{Conformally invariant $\left<\phi^{2}\right>$}
\label{cphi2}
It is of interest to present $\left<\phi^{2}\right>$ in eq. (\ref{termsphi2}) for a conformally invariant scalar field ($M=0$ and $\xi=1/6$). This can be easily done by ensembling the results above.
It follows that
\begin{eqnarray}
	&&
	\left<\phi^{2}\right>=-\frac{1}{48\pi^2 a^{2}}+
	\frac{1}{2\pi^2 a^{2}}e^{-1/Ta}+\cdots, \hspace{1.0cm} Ta\ll 1
	\label{phi2-1}
	\\
	&&
	\left<\phi^{2}\right>=\frac{T^2}{12}-\frac{T}{4\pi^2 a}
	-2T^2 e^{-4\pi^2 Ta}+\cdots, \hspace{0.9cm} Ta\gg 1
	\label{phi2-2}
\end{eqnarray}
where eqs. (\ref{vphi2-2}), (\ref{tphi2-2}), (\ref{mphi2-4}) and (\ref{mphi2-5}) have been used. It is worth remarking that blackbody's contribution in eq. (\ref{phi2-2}) is corrected by a term that arises due to the finite size of the universe. By summing first over $m$ instead,
such a term would be lost. Correspondingly the expression in eq. (\ref{phi2-1}) would gain the contribution $T/4\pi^2 a$.

\subsection{$\xi$-dependent corrections}
\label{corrections-phi2}
Before closing the study of $\left<\phi^{2}\right>$, let us discuss briefly the issue of obtaining $\xi$-dependent corrections to leading contributions  
since that might be of some interest
\footnote{In fact, $\xi$-dependent corrections in $\left<T^\mu{}_\nu\right>$ will be relevant in the study of thermodynamic stability.}. As an illustration we will address the massless $\left<\phi^{2}\right>_{{\tt thermal}}$ with an assumption which will be very much used along the text, namely
\begin{equation}
	M=0, \hspace{0.5cm} \frac{\sqrt{6\xi-1}}{Ta}\rightarrow 0	
	\label{assumption1}	
\end{equation} 
A word of caution in applying eq. (\ref{assumption1})
is in order. We notice that $Ta\ll 1$ can perfectly well be considered as long as $\xi$ is close enough to $1/6$ such that the asymptotic regime in eq. (\ref{assumption1}) is not spoiled.

\begin{figure}[htbp]
	\centering
	\includegraphics[scale=1]{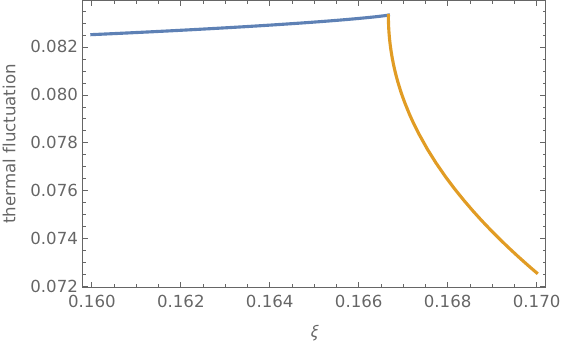}
	\caption{\footnotesize{Massless $\left<\phi^{2}\right>_{{\tt thermal}}$  around $\xi =1/6$, for unitary $a$ and $T$. On the left (blue) and on the right (yellow) are the plots corresponding to eqs. (\ref{tphi2-5}) and (\ref{tphi2-6}), respectively.}}
	\label{fig3}
\end{figure}

The technicalities involved are identical to those in ref. \cite{med01} to calculate mass corrections to
thermal fluctuations in flat backgrounds.
Considering the integral representation for $K_{\nu}({\rm z})$ in eq. (9.6.23) of ref. \cite{abr65}, and after 
some manipulations \cite{med01}, it can be shown that
\begin{eqnarray}
	&&
	\left<\phi^{2}\right>_{{\tt thermal}}=\frac{T^2}{12}+
	\frac{1-6\xi}{8\pi^2 a^2}\left(\ln	\frac{\sqrt{1-6\xi}}{Ta}+\cdots \right)+ \cdots, \hspace{2.7cm} \xi<\frac{1}{6}
	\label{tphi2-5}
	\\
	&&
	\left<\phi^{2}\right>_{{\tt thermal}}=\frac{T^2}{12}-
	\frac{\sqrt{6\xi-1}}{4\pi a}T+\frac{1-6\xi}{8\pi^2 a^2}\left(\ln	\frac{\sqrt{6\xi-1}}{Ta}+\cdots \right)+\cdots, \hspace{0.5cm} \xi>\frac{1}{6}.
	\label{tphi2-6}
\end{eqnarray}
Note that the linear term on $T$ is missing in eq. (\ref{tphi2-5})
[cf. Figure \ref{fig3}].

\section{$\left<T^\mu{}_\nu\right>$}
\label{stress}

We turn now to the ensemble average of the stress-energy-momentum tensor
at finite temperature in the EU. Here also we shall use the the renormalized $G_{E}$ in eq. (\ref{rfeynman}). Formally, 
\begin{eqnarray}
	&&
	\left<T_\mu{}_\nu\right>=
	\Re\left(
	\lim_{{\rm x}'\rightarrow{\rm x}} D_{\mu{}\nu}({\rm x},{\rm x}')
	G({\rm x},{\rm x}')\right)
	\label{defstress}
	\\
	&&
	\hspace{1.04cm}
	=\left<T_\mu{}_\nu\right>_{{\tt vacuum}}+
	\left<T_\mu{}_\nu\right>_{{\tt mixed}}+
	\left<T_\mu{}_\nu\right>_{{\tt thermal}}
	\label{termsstress}	
\end{eqnarray}
where $D_{\mu{}\nu}({\rm x},{\rm x}')$ is a differential operator acting on the biscalar $G({\rm x},{\rm x}')$ given by
\begin{equation}
	D_{\mu \nu}({\rm x},{\rm x}'): = (1-2\xi)\nabla_\mu\nabla_{\nu'}+(2\xi -\frac{1}{2})g_{\mu{}\nu}\left(g^{\rho{}\sigma}\nabla_\rho\nabla_{\sigma'}-M^2-\xi R\right)-2\xi\nabla_\mu\nabla_{\nu}-\xi R_{\mu\nu}.
	\label{operator}
\end{equation}
The use of eq. (\ref{operator}) in eq. (\ref{defstress}) yields 
$\left<T_\mu{}_\nu\right>$ corresponding to
the classical $T_\mu{}_\nu$ in eq. (3.190) of ref. \cite{dav82}. Once the covariant derivatives have been evaluated (see appendix \ref{geometry}) in eq. (\ref{defstress}), one proceeds along the same lines as in the calculation of $\left<\phi^{2}\right>$ in the previous section to obtain each contribution in eq. (\ref{termsstress}). It results that 
$\left<T_\mu{}_\nu\right>$ 
is diagonal
\footnote{Note that, of course, $\left<T^\mu{}_\nu\right>=g^{\mu{}\alpha}\left<T_\alpha{}_\nu\right>$.},
i.e., 
\begin{eqnarray}
\left<T^\mu{}_\nu\right>={\rm diag}	(\rho,-p,-p,-p).
\label{density-pressure}	
\end{eqnarray}
Moreover
\begin{eqnarray}
	p=\frac{\rho}{3}-\frac{M^2}{3}\left<\phi^{2}\right>.
	\label{eq-of-state}	
\end{eqnarray}
Thus all properties of $\left<T^\mu{}_\nu\right>$ can be extracted from
$\rho$ and $\left<\phi^{2}\right>$. For instance, a glance at eqs. 
(\ref{density-pressure}) and (\ref{eq-of-state}) yields its trace,
 $$\left<T^\mu{}_\mu\right>=M^2\left<\phi^{2}\right>$$
It can also be shown that $\nabla_\mu\left<T^\mu{}_\nu\right>=0$.

\subsection{Energy density}
\label{energydensity}
Likewise $\left<\phi^{2}\right>$ in eq. (\ref{termsphi2}), the energy density in eq. (\ref{density-pressure}) is given by the sum of three typical contributions, namely 
\footnote{Consequently, considering eq. (\ref{eq-of-state}), the same applies to the pressure, i.e.,
$p=p_{{\tt vacuum}}+p_{{\tt mixed}}+p_{{\tt thermal}}$.}
\begin{eqnarray}
	\rho
	=\rho_{{\tt vacuum}}+
	\rho_{{\tt mixed}}+
	\rho_{{\tt thermal}} 
	&&
	\label{termsenergydensity}	
\end{eqnarray}
We notice that the comments following eq. (\ref{termsphi2}) still apply here.

\subsubsection{$\rho_{{\tt vacuum}}$}
\label{ved}

One can show that
\begin{eqnarray}
	&&	 \rho_{{\tt vacuum}}= 
	\frac{1}{4\pi^3a^4}\Re\Bigg(\sum_{n = 1}^{\infty}\bigg[
	\frac{(M^2 a^2+6 \xi -1)^{3/2}}{n}K_1\left(2\pi n\sqrt{M^2 a^2+6 \xi -1}\right)
	\hspace{2.0cm}
	\nonumber
	\\ 
	&&\hspace{4.0cm} +3\,\frac{M^2 a^2+6 \xi -1}{2\pi n^2}K_2\left(2\pi n  \sqrt{M^2 a^2+6 \xi -1}\right)\bigg]\Bigg)
	\label{vrho1}
\end{eqnarray}
If $M^2 a^2+6 \xi -1$ is nonnegative then we do not have to bother with $\Re$ in eq. (\ref{vrho1}), resulting precisely eq. (12) of ref. \cite{her06}
for the vacuum energy density in the EU
where a well different renormalization procedure 
has been applied
\footnote{By setting $\xi=1/6$, eq. (\ref{vrho1}) also reproduces eq. (10) in ref. \cite{dow77}, apart a couple of typos.}. 

\begin{figure}[htbp]
	\centering
    \includegraphics[scale=0.8]{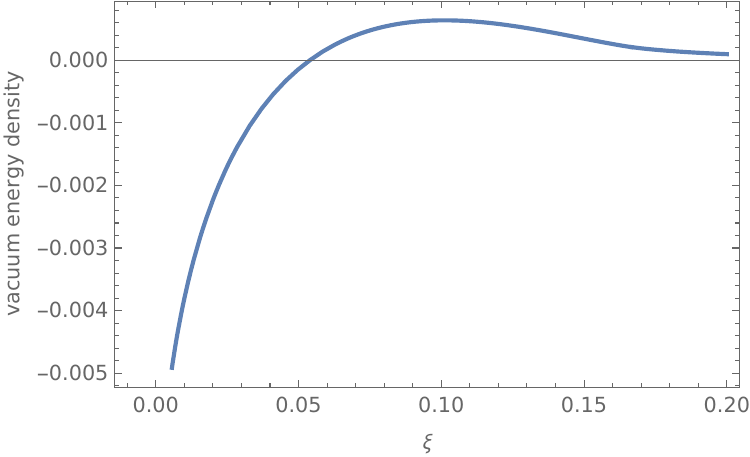}
	\caption{\footnotesize{$\rho_{{\tt vacuum}}$ vs. $\xi$, for $\xi>0$, $M=0$ and $a=1$. The vacuum energy density vanishes at $\xi\simeq 0.054$, changes sign there, and as $\xi\rightarrow\infty$ it approaches zero from above.}}
	\label{fig4}
\end{figure}

\begin{figure}[htbp]
	\centering
	\includegraphics[scale=0.8]{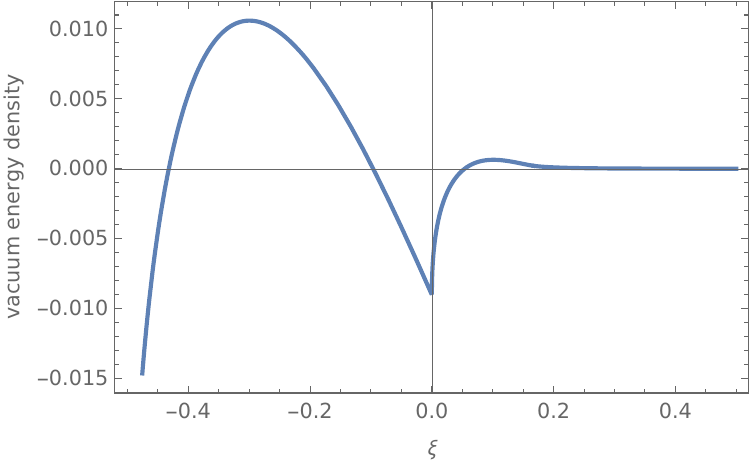}
	\caption{\footnotesize{$\rho_{{\tt vacuum}}$ vs. $\xi$, for $\xi>-1/2$, $M=0$ and $a=1$. At $\xi=0$,  $\rho_{{\tt vacuum}}\simeq -0.009$. The cusp-like pattern showing in the plot repeats with increasing amplitude as $\xi\rightarrow -\infty$.}}
	\label{fig5}
\end{figure}

By using again eq. (\ref{ab1}) and that
$\zeta(4)=\pi^4/90$, it follows 
\footnote{Massive fields can be addressed simply by adding $M^2a^2$ to $6\xi-1$ in the formulae which holds as long as $M^2a^2+6\xi-1\ll 1$.}
\begin{equation}
	\rho_{{\tt vacuum}}= \frac{1}{480\pi^2 a^4}
	[1-5(6\xi-1)+\cdots],
	\hspace{3.0cm} 
	M=0, \hspace{0.2cm} \xi\rightarrow\frac{1}{6}
	\label{vrho2}
\end{equation}
which reproduces the well known result in eq. (\ref{cved}) when $\xi=1/6$
(conformal coupling), and it agrees with ref. \cite{eli03} where $\zeta$-function regularization has been used.
The plots in Figures \ref{fig4} and \ref{fig5} show a rather unsuspected behavior. Namely, although massless $\rho_{{\tt vacuum}}$ is positive at $\xi=1/6$ [cf. eq. (\ref{vrho2})], as $\xi$ decreases from $1/6$, $\rho_{{\tt vacuum}}$ changes sign turning out to be negative when $\xi=0$ (minimal coupling).
Such a behavior is in agreement with the study in ref. \cite{her06}, but it is at variance with refs. \cite{zhu96,alt03} 
\footnote{Ref. \cite{zhu96} claims that $\rho_{{\tt vacuum}}>0$ for $\xi\in [0,1/6]$ whereas ref. \cite{alt03} states that  $\rho_{{\tt vacuum}}=0$ when $\xi=0$. 
}.


\subsubsection{$\rho_{{\tt thermal}}$}
\label{ted}
Now we turn to $\rho_{{\tt thermal}}$ in eq. (\ref{termsenergydensity})
which is given by
\begin{eqnarray}
	&&	 \rho_{{\tt thermal}}= 
	\frac{T}{2\pi^2a^3}\Re\Bigg(\sum_{m = 1}^{\infty}\bigg[
	\frac{(M^2 a^2+6 \xi -1)^{3/2}}{m}
	K_1\left(\frac{m\sqrt{M^2 a^2+6 \xi -1}}{Ta}\right)
	\hspace{2.0cm}
	\nonumber
	\\ 
	&&\hspace{4.0cm} +3Ta\,\frac{M^2 a^2+6 \xi -1}{m^2}
	K_2\left(\frac{m\sqrt{M^2 a^2+6 \xi -1}}{Ta}\right)\bigg]\Bigg).
	\label{trho1}
\end{eqnarray}
Just as for $\left<\phi^{2}\right>_{{\tt thermal}}$, when $\xi=1/6$,  eq. (\ref{trho1}) becomes the blackbody energy density in Minkowski spacetime (see e.g. ref. \cite{kap06}), losing therefore its dependence on $a$. 
Another feature worth remarking on is that by setting $1/T$ equal to $2\pi a$
one goes from $\rho_{{\tt thermal}}$ in eq. (\ref{trho1})
to $\rho_{{\tt vacuum}}$ in eq. (\ref{vrho1}) \footnote{This is typical of thermodynamics of quantum fields in backgrounds with nontrivial topology.}.

Assuming eq. (\ref{assumption1}),
the asymptotic expressions in eq. (\ref{ab1}) yields
	\begin{equation}
	\rho_{{\tt thermal}}= \frac{\pi^2 T^{4}}{30}-\frac{6\xi-1}{24 a^{2}} T^2+\cdots .
	\label{trho2}
\end{equation}
 We notice that, unlike $\left<\phi^{2}\right>_{{\tt thermal}}$ in eqs. (\ref{tphi2-5}) and (\ref{tphi2-6}), there is no cusp at $\xi=1/6$ [cf. Figure \ref{fig3}]
corresponding to $\rho_{{\tt thermal}}$.
Although the term carrying $\xi$ and $a$ in eq. (\ref{trho2}) is a subleading contribution to the dominant blackbody term, we should stress the relevance of its correctness  which will be crucial for a reliable analysis of thermodynamic stability in the next sections.

\begin{figure}[htbp]
	\centering
	\includegraphics[scale=0.8]{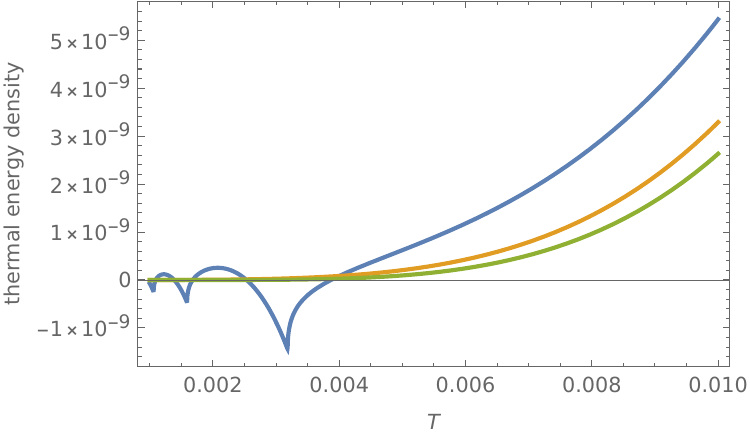}
	\caption{\footnotesize{$\rho_{{\tt thermal}}$ vs. $T$, for $M=0$ and $a=1$. On the right: the lower (green) plot,
			the middle (yellow) plot, and the upper 	(blue) plot correspond,
			respectively, to $\xi=0.1667$, $\xi=1/6$,
		    and $\xi=0.1666$. Again, note that these asymptotic behaviors
			as $T\rightarrow 0$ differ radically from each other in spite of their  associated values of $\xi$ are not that different.}}
	\label{fig6}
\end{figure}
As anticipated previously in the text, for $\xi<1/6$, eq. (\ref{trho1}) shows that $\rho_{{\tt thermal}}$ also presents an oscillatory cusp-like behavior as $Ta\rightarrow 0$
which will raise thermodynamic stability issues in the next section. For $\xi>1/6$,  $\rho_{{\tt thermal}}$ presents an exponential decay as $Ta\rightarrow 0$. And For $\xi=1/6$, eq. (\ref{trho2}) holds.
(cf. Figure \ref{fig6}).

\subsubsection{$\rho_{{\tt mixed}}$}
\label{med}
We now present the last contribution in eq. (\ref{termsenergydensity}),
namely
\begin{eqnarray}
	&&	 \rho_{{\tt mixed}}= 
	(M^2 a^2+6 \xi -1)\frac{T}{\pi^2a^2}\Re\Bigg(\sum_{m,n = 1}^{\infty}
	\Bigg[\sqrt{M^2 a^2+6 \xi -1}\,
	\frac{(2\pi n Ta)^{4}-(4\pi m n Ta)^{2}+m^{4}}{a[(2\pi n Ta)^{2}+m^{2}]^{5/2}}\hspace{0cm}
	\nonumber
	\\
	\nonumber
	&& \hspace{6.0cm}\times K_1\left(\frac{\sqrt{M^2 a^2+6 \xi -1}}{Ta}\sqrt{(2\pi n Ta)^{2}+m^{2}}\right)
	\\ 
	&&\hspace{1.8cm} 
    +T\left(3\frac{(2\pi n Ta)^{4}-6(2\pi m n Ta)^{2}+m^{4}}{[(2\pi n Ta)^{2}+m^{2}]^{3}}
	-\frac{\left(M^2 a^2+6 \xi -1\right)(2\pi m n)^{2}}{[(2\pi n Ta)^{2}+m^{2}]^2}\right)
	\nonumber
	\\
	&& \hspace{6.0cm}\times K_2\left(\frac{ \sqrt{M^2 a^2+6 \xi -1}}{Ta}\sqrt{(2\pi n Ta)^{2}+m^{2}}\right)\Bigg]\Bigg).
	\label{mrho-1}
\end{eqnarray}
By applying eq. (\ref{ab1}) considering eq. (\ref{assumption1}), it follows that
\begin{eqnarray}
	&&	 \rho_{{\tt mixed}}= 
	\frac{6T^4}{\pi^2}\sum_{m,n=1}^{\infty}
	\frac{(2\pi n Ta)^{4}-6(2\pi m n Ta)^{2}+m^{4}}{[(2\pi n Ta)^{2}+m^{2}]^{4}}\hspace{5.0cm}
	\nonumber
	\\
	&&
	\hspace{3.0cm}
	-\frac{(6\xi-1)T^2}{2\pi^2 a^2}\sum_{m,n=1}^{\infty}
	\frac{(2\pi n Ta)^{4}-6(2\pi m n Ta)^{2}+m^{4}}{[(2\pi n Ta)^{2}+m^{2}]^{3}}+\cdots.
	\label{mrho-2}	
\end{eqnarray}

The double series above have the same nature of that in 
eq. (\ref{mphi2-2}) with a remark that now the order in which the summations are evaluated does not matter. In fact, we can proceed exactly as we did to obtain eqs. (\ref{mphi2-4}) and (\ref{mphi2-5}). In doing so we end up with,
\begin{eqnarray}
	&&
	\rho_{{\tt mixed}}=-\frac{\pi^2 T^{4}}{30}+\frac{6\xi-1}{24 a^{2}} T^2+
	\frac{1}{2\pi^2 a^{4}}[1+\cdots] e^{-1/Ta}+\cdots, \hspace{0.82cm} Ta\ll 1
	\label{mrho-3}
	\\
	&&
	\rho_{{\tt mixed}}=-\frac{1}{480\pi^2 a^4}+\frac{6\xi-1}{96 \pi^2 a^{4}}
	+
	8\pi^2 T^4 [1+\cdots]e^{-4\pi^2 Ta}+\cdots, \hspace{0.5cm} Ta\gg 1.
	\label{mrho-4}
\end{eqnarray}
Since these expressions will be used shortly, 
we notice that the first two terms in eq. (\ref{mrho-3}) are
$-\rho_{{\tt thermal}}$ in eq. (\ref{trho2}), and that the first two terms in eq. (\ref{mrho-4}) are $-\rho_{{\tt vacuum}}$ in eq. (\ref{vrho2}).

Here also the whole story repeats itself when, for a given $\xi$, $Ta\rightarrow 0$. 
If $\xi<1/6$, $\rho_{{\tt mixed}}$ develops
an oscillatory cusp-like behavior with a pattern different from that of $\rho_{{\tt thermal}}$
\footnote{Resulting that  $\rho$ in eq. (\ref{termsenergydensity}) also shows an oscillatory cusp-like behavior when $Ta\rightarrow 0$ and $\xi<1/6$. Of course, such a pattern vanishes when one sets $T=0$.}. 
If $\xi>1/6$, one has 
a $-\rho_{{\tt mixed}}$ that decays
exponentially as $Ta\rightarrow 0$. If $\xi=1/6$, $\rho_{{\tt mixed}}$ 
behaves according eq. (\ref{mrho-3}).
\subsubsection{Conformally invariant $\rho$}
\label{crho}
In order to check consistency with the literature, let us derive the energy density of a conformally invariant scalar field.
Noticing eq. (\ref{termsenergydensity}) and 
setting $\xi=1/6$ in eqs. (\ref{vrho2}), (\ref{trho2}), (\ref{mrho-3}) and (\ref{mrho-4}), we are led to
\begin{eqnarray}
	&&
	\rho=\frac{1}{480\pi^2 a^4}
	+
	\frac{1}{2\pi^2 a^{4}}e^{-1/Ta}+\cdots, \hspace{1.0cm} Ta\ll 1
	\label{rho-1}
	\\
	&&
	\rho=\frac{\pi^2 T^{4}}{30}
	+8\pi^2 T^4 e^{-4\pi^2 Ta}+\cdots, \hspace{1.4cm} Ta\gg 1
	\label{rho-2}
\end{eqnarray}
which are well known expressions that have been discovery in the early papers on quantum fields in the EU (see section \ref{introduction}). It should be pointed out that the blackbody term in eq. (\ref{rho-2}) is corrected  only by an exponentially small term containing the size of the universe. Nevertheless, we stress, this is not the case when $\xi\neq 1/6$ 
where the contribution due to the finite size of the EU is much larger [cf. eq. (\ref{trho2})], a fact that will be relevant in the study of thermodynamic stability as has been already mentioned in the text.
Perhaps 
it is worth remarking that eq. (\ref{rho-1}) as well as its counterpart eq. (\ref{phi2-1}) still hold when $Ta\rightarrow 0$.

\subsection{Remarks on statistical mechanics}
\label{statmech}

As explained in section \ref{introduction}, one can also apply a global approach to reach the blackbody's thermodynamics by considering the statistical mechanics of an ideal gas of bosons with zero chemical potential in the EU [cf. eq. (\ref{sm1}) and its paragraph]. In order to address the nonminimal coupling in eq. (\ref{wave2}), we consider the following energy spectrum \cite{her06},
\begin{equation}
	\epsilon_{n} = \sqrt{\frac{(n+1)^2+6\xi-1}{a^{2}}+ M^2 },
	\hspace{1.0cm}
	n=0,1,2,\cdots,
	\hspace{1.0cm}
	g_{n}=(n+1)^{2}
	\label{spectrum2}
\end{equation}
Clearly, in this approach, only non negative values of $\xi$ are allowed.

 By feeding eq. (\ref{sm1}) with eq. (\ref{spectrum2}), it results
 \cite{eli03}
\begin{equation}
\rho_{T}=\frac{1}{2\pi^2 a^{4}}\sum_{l=1}^{\infty}
\frac{l^2\sqrt{l^2+M^2 a^2+6\xi-1}}{e^{\frac{1}{Ta}\sqrt{l^2+M^2 a^2+6\xi-1}}-1}
\label{rhot1}	
\end{equation}
which, with help of Abel-Plana's formula, yields
\footnote{We notice that a factor $\pi^2/6$ is missing in the term containing $\beta^{-2}$ in eq. (54) of ref. \cite{eli03}.}
	\begin{equation}
	\rho_{{T}}= \frac{\pi^2 T^{4}}{30}-\frac{M^2 a^2+6\xi-1}{24 a^{2}} T^2+\cdots, \hspace{1.0cm} Ta\rightarrow \infty
    \label{rhot}
\end{equation}
and thus in agreement with the dominant contributions resulting from the sum of $\rho_{{\tt thermal}}$ in eq. (\ref{trho2}) with $\rho_{{\tt mixed}}$ in eq. (\ref{mrho-4}). So far no surprises.

In order to study  $\rho_{T}$ as $Ta\rightarrow 0$, we go back to eq. (\ref{rhot1}). We notice that for massless bosons with $\xi=0$ the term corresponding to $l=1$ in the summation produces a ``zero-mode'' divergence about which we have more to say shortly (see ref. \cite{mor20}, and references therein).
For $\xi>0$ \footnote{Or $\xi=0$ and $M>0$.}, one has that
\begin{equation}
	\rho_{T}=\frac{\sqrt{M^2 a^2+6\xi}}{2\pi^2 a^{4}}
	e^{-\sqrt{M^2 a^2+6\xi}/Ta}+\cdots, \hspace{0.6cm} Ta\rightarrow 0
	\label{rhot2}	
\end{equation}
which is eq. (25) in ref. \cite{eli03}.
In particular for massless scalar bosons,
\begin{equation}
	\rho_{T}=\frac{\sqrt{6\xi}}{2\pi^2 a^{4}}
	e^{-\sqrt{6\xi}/Ta}+\cdots, \hspace{2.9cm} Ta\rightarrow 0
	\label{rhot3}	
\end{equation}
therefore contrasting radically with $\rho_{{\tt thermal}}+\rho_{{\tt mixed}}$ in the local approach which, as we know,  acquires a typical oscillatory cusp-like behavior
when $\xi<1/6$ and $Ta\rightarrow 0$.
Nevertheless, when we set $\xi=1/6$ in eq. (\ref{rhot3}) and add
$\rho_{{\tt vacuum}}$, we do reproduce eq. (\ref{rho-1}) of the local approach.
At this point it is worth clarifying  that, for a given $a$, both 	$\rho_{T}$
and $\rho_{{\tt thermal}}+\rho_{{\tt mixed}}$ behave as in eq. (\ref{rhot}) at high temperatures, and only departure
from each other when $T\rightarrow 0$ for $\xi<1/6$.

Now to deal with $\xi=0$ and $M=0$, we may set $\xi=0$ with non vanishing $M$ in eq. (\ref{rhot1}), following by taking $M\rightarrow 0$, yielding
\begin{equation}
	\rho_{T}=\frac{T}{2\pi^2 a^3}+\cdots, \hspace{4.5cm} Ta\rightarrow 0.
	\label{rhot4}	
\end{equation}
We should remark that, as $T\rightarrow 0$ for a given $a$, massless 
$\rho_{T}$ with $\xi\neq 0$, but otherwise very close to zero, behaves initially according eq. (\ref{rhot4}), and then according to  eq. (\ref{rhot3}).

\section{Thermodynamic stability}
\label{stability}

Finally we get to the main topic in this work. Namely, the study on how $\xi$ is restricted by imposing criteria of stable thermodynamic equilibrium to the massless scalar blackbody radiation in the EU.
As an application, thermodynamic stability of a mixture of radiations will be explored in the light of the results.

\subsection{Permissible values of $\xi$}
\label{pvalues}
Let us anticipate by saying that, strictly speaking, the only value of $\xi$ for a massless scalar radiation that complies with the criteria for thermodynamic stability at all $T$ and for all $a$ in the EU is that corresponding to conformal coupling, i.e., $\xi=1/6$. 
More precisely, for a given $a$, we shall see that $\xi\neq 1/6$
results in thermodynamic instabilities in the EU at low and high temperatures.

The criteria implying thermal and mechanical stability are, respectively \cite{cal85},
\begin{equation}
	C_{V}:=\left(\frac{\partial U}{\partial T}\right)_{V}>0, \hspace{2.0cm} 
	\kappa_{T}:=-\frac{1}{V}\left(\frac{\partial V}{\partial p}\right)_{T}>0
	\label{criteria1}
\end{equation}
where $C_{V}$ is the heat capacity at constant volume $V$ and
$\kappa_{T}$ is the isothermal compressibility. Therefore, both $C_{V}$ and  $\kappa_{T}$ must be positive for a system at stable thermodynamic equilibrium. Noticing eqs. (\ref{eq-of-state}) and (\ref{volume}) and recalling that $U=\rho V$, it follows that the inequalities in eq. (\ref{criteria1}) for radiation in the EU become
\begin{equation}
	\left(\frac{\partial \rho}{\partial T}\right)_{a}>0, \hspace{2.0cm} 
	\left(\frac{\partial \rho}{\partial a}\right)_{T}<0.
	\label{criteria2}
\end{equation}
That is, the energy density in eq. (\ref{termsenergydensity}) must be a increasing function of $T$, and a decreasing function of $a$.

\subsubsection{$Ta\rightarrow 0$}
\label{vlowt}
At this regime, when $\xi<1/6$,  $\rho$ vs. $T$ in eq. (\ref{termsenergydensity}) acquires 
an oscillatory cusp-like pattern around ${\rho_ {\tt vacuum}}$
[cf. Figure \ref{fig6}], resulting in numerous regions where $\partial_{T}\rho$ is negative, and thus violating thermal stability. Moreover, at $T$ corresponding to a cusp, $\partial_{T}\rho$ is ill-defined, i.e., $C_{V}$ is ill-defined.
Note also that around $\xi=0$ and other regions with $\xi<0$, ${\rho_ {\tt vacuum}}<0$ [see Figure \ref{fig5}]. Then, as ${|\rho_ {\tt vacuum}}|$ drops with $a$ [cf. eq. (\ref{vrho1})],
one gets that $\partial_{a}\rho>0$ violating mechanical stability.
We then conclude that when $\xi<1/6$ the scalar blackbody radiation becomes thermodynamically unstable as $Ta\rightarrow 0$. On the other hand,
when $\xi\geq 1/6$ no such pathologies affect $\rho$ resulting that eq. (\ref{criteria2}) holds as $Ta\rightarrow 0$.

\subsubsection{$ \sqrt{6\xi-1}/Ta\rightarrow 0$}
\label{lht}
As mentioned previously, there are two situations in which 
the asymptotic regime in eq. (\ref{assumption1})  takes place.
\vspace{0.1cm}

{\it $Ta\ll 1$}

\noindent This can be considered as long as $\xi\rightarrow 1/6$. Thus eq. (\ref{rho-1}) holds approximately and both inequalities in eq. (\ref{criteria2}) are satisfied ensuring thermodynamic stability.

\vspace{0.1cm}
{\it $Ta\gg 1$}

\noindent This can be considered for any value of $\xi$ when $Ta\rightarrow \infty$. Now, looking at eq. (\ref{termsenergydensity})
and considering eqs. (\ref{trho2}) and (\ref{mrho-4}), it results
\begin{equation}
	\rho= \frac{\pi^2 T^{4}}{30}-\frac{6\xi-1}{24 a^{2}} T^2+\cdots, \hspace{1.0cm} Ta\rightarrow \infty.
	\label{rho3}
\end{equation}
We notice that eq. (\ref{rho3}) for $\xi=1/6$ yields eq. (\ref{rho-2})
which clearly meets  the criteria of thermodynamic stability in eq. (\ref{criteria2}). When $\xi<1/6$, one can promptly see that eq. (\ref{rho3})  also leads to eq. (\ref{criteria2}). In contrast, for  $\xi>1/6$, it results that
$\partial_{a}\rho>0$, i.e., violation of mechanical stability.
The conclusion is that when $\xi>1/6$ the scalar blackbody radiation
in the EU becomes thermodynamically unstable as $Ta\rightarrow \infty$.

\subsection{Mixed particles hot soup}
\label{application}
As an exercise on thermodynamic stability, consider three species of blackbody radiations filling the EU at the regime $Ta\rightarrow \infty$.
The mixture contains $n_{\mu}$ scalar fields with the same curvature coupling parameter $\xi$, $n_{\nu}$ spinor fields, and $n_{\gamma}$ vector fields. All fields are assumed to be massless and thus $p=\rho/3$, where
\begin{equation}
	\rho=n_{\mu}\rho_{\mu}+n_{\nu}\rho_{\nu}+n_{\gamma}\rho_{\gamma}
\label{rho4}	
\end{equation}
with  $\rho_{\mu}$ as in eq. (\ref{rho3}), and the other blackbody energy densities given by \cite{alt78}
\begin{equation}
	\rho_{\nu}= \frac{7\pi^2 T^{4}}{60}-\frac{T^2}{24 a^{2}}+\cdots, \hspace{1.0cm} 
	\rho_{\gamma}= \frac{\pi^2 T^{4}}{15}-\frac{T^2}{6 a^{2}}+\cdots
	\label{rho5}
\end{equation}
Since $\partial_{T}\rho>0$, thermal stability holds [cf. eq. (\ref{criteria2})].

Now, regarding mechanical stability, we have that
\begin{equation}
	\left(\frac{\partial \rho}{\partial a}\right)_{T}=\frac{T^2}{12 a^3}\left[(6\xi-1)n_{\mu}+n_{\nu}+4n_{\gamma}\right]
\label{dda}	
\end{equation}
As $\partial_{a}\rho$ in eq. (\ref{dda}) must be negative to meet the criterion of mechanical stability [cf. eq. (\ref{criteria2})], when spinors and/or vectors are present at least one scalar field must be present, with $\xi$ satisfying
\begin{equation}
	\xi<\frac{1}{6}\left(1-\frac{n_{\nu}+4n_{\gamma}}{n_{\mu}}\right).
\label{bound}	
\end{equation}
It follows then the curious fact that, for a given radius $a$, neutrinos or photons cannot be on their own in a hot EU if thermodynamic stability is required
\footnote{In contrast, by using formulae in ref. \cite{alt78}, one may check that electromagnetic and neutrino blackbody radiations in the EU are thermodynamic stable at the regime $Ta\rightarrow 0$.}.

\section{Final remarks}
\label{conclusion}

In this paper we have derived the renormalized Feynman propagator, at finite temperature $T$, for a massive and neutral scalar field $\phi$ in the EU of radius $a$, for arbitrary real values of  the curvature coupling parameter $\xi$.
A thorough analysis on how the ensemble averages $\left<\phi^{2}\right>$ and 
$\left<T^\mu{}_\nu\right>$ depend on $T$ and $\xi$ has been presented,
revealing  that when $Ta\rightarrow 0$ only $\xi\geq 1/6$ are allowed
by thermodynamic stability of a massless $\phi$.
It follows that the minimal coupling, $\xi=0$, should be discarded at low temperatures and/or small radius of the EU.
 Similarly, when $Ta\rightarrow \infty$,
only $\xi\leq 1/6$ meets the criteria of thermodynamic stability of the massless scalar blackbody radiation in the EU.
We have shown a discrepancy between statistical mechanics and the local quantum field approach for $\xi\neq 1/6$, which seems to reflect the non equivalence of global equilibrium and a local operator based notion of thermodynamic equilibrium in compact and curved spacetimes.
We then applied our findings to a mixture of scalar bosons,
neutrinos and photons showing that at least one scalar field must be present such that the hot atmosphere is thermodynamically stable
with $\xi$ tuned as in eq. (\ref{bound}). Therefore one concludes that neutrinos as well as photons cannot be alone in the EU when $Ta\rightarrow \infty$, which sounds like a surprise.

An early paper by Hu (see ref. \cite{hu82}, and references therein) has explored thermodynamic equilibrium in non stationary universes
\footnote{We wish to report two apparent typos in ref. \cite{hu82}.
To be consistent with ref. \cite{wel81}, the term $m^2 T^2/24 a^2$
in eq. (8)
should be replaced by  $-m^2 T^2/24$. And the factor $T^2/24$ in eq. (10), which is a generalization of eq. (\ref{rhot}) in the present account, should be $-T^2/24$ instead. These points do not affect the conclusions in ref. \cite{hu82}.}.
In particular for massless $\phi$ in Robertson-Walker universes, it has been shown that only $\xi=1/6$ is consistent with the notion of thermodynamic equilibrium holding at all times. The author then introduced the notion of quasi-thermodynamic equilibrium, according to which $\xi>1/6$ would also be allowed as long as $(6\xi-1)R/6\ll T^2$, where $R$ is the Ricci scalar curvature. Now, thermodynamics is also 
about stable thermodynamic equilibrium \cite{cal85}. Thus, it remains to be seen whether $\xi>1/6$ at high temperatures would not also be discarded by thermodynamic stability in Robertson-Walker universes. This point needs careful investigation beyond the scope of our paper.

We wish now to indulge in a speculation regarding the behavior of $\left<\phi^{2}\right>$ and $\left<T^\mu{}_\nu\right>$ when $Ta\rightarrow 0$ and $\xi<1/6$. As we know well by now, these averages will present oscillatory cusp-like patterns that could be eliminated by simply setting $\xi\geq 1/6$, in which case their physical meaning would be just an indication of the breakdown of thermodynamic consistency. Nevertheless, one might wonder if these pathological patterns are not telling us more than that we cannot have $\xi<1/6$ near the absolute zero of temperature and/or $a\rightarrow 0$. Perhaps this point also deserves further investigation.

A point worth clarifying concerns a possible relation between the thermodynamic instability discussed here and the well-known dynamical instability of the Einstein universe as a solution of Einstein's equations. The latter arises from the gravitational dynamics of the scale factor and reflects the fact that the static configuration corresponds to an unstable equilibrium under the interplay between matter and the cosmological constant. In contrast, the present analysis is performed on a fixed $R\times S^3$
background and addresses the thermodynamic stability of quantum fields as  characterized by positivity of the heat capacity and of the isothermal compressibility.
These two notions of stability are therefore conceptually distinct. Nevertheless, they are not entirely unrelated. In the present context, the Einstein universe is supported by classical sources (dust and a positive cosmological constant), and the expectation value $\left<T^\mu{}_\nu\right>$ provides an additional quantum contribution to the total stress-energy-momentum  tensor. When backreaction is taken into account, this quantum contribution would lead to a perturbative modification of the background geometry. Since $\left<T^\mu{}_\nu\right>$ 
depends on the curvature coupling parameter 
$\xi$, it may in principle influence the dynamical stability of the configuration, either enhancing or mitigating the classical instability. A detailed analysis of this interplay, however, lies beyond the scope of the present work.

A last word before closing. The essence of this paper is theoretical, of course, and focuses on a particular static universe. Nevertheless, thermodynamic stability is an issue that should be addressed in any model of the early universe.
We expect the findings here to be useful in this scenario.

\appendix
\section{Appendixes}
The following material, which involves technicalities, complements the main text. 

\subsection{Geometric objects}
\label{geometry}

Considering eq. (\ref{le}), it is straightforward to obtain the following objects in  $R\times S^{3}$. Our conventions are those in ref. \cite{dav82}.

When $S^{3}$ is taken as immersed in $R^{4}$, satisfying $X^2+Y^2+Z^2+W^2=a^{2}$,
the angle $\alpha$ between $V=(X,Y,Z,W)$ and $V'=(X',Y',Z',W')$ on $S^{3}$ is given by
$\alpha=\arccos\left(V\cdot V'/a^2\right)$
and the corresponding geodesic distance is $s=a\alpha$.
By using spherical polar coordinates [see eq. (\ref{le})], 
\begin{eqnarray}
X=a\sin\chi\sin\theta\cos\varphi, \hspace{0.5cm}
Y=a\sin\chi\sin\theta\sin\varphi, \hspace{0.5cm}	
Z=a\sin\chi\cos\theta, \hspace{0.5cm}	
W=a\cos\chi 
&&
\nonumber
\end{eqnarray}
one finds that
\begin{equation}
\alpha = \arccos \left\{\cos\chi \cos\chi'+\sin\chi \sin\chi'\left[\cos\theta \cos\theta'+ \sin\theta \sin\theta'\cos(\varphi-\varphi')\right]\right\}.
\label{angle}
\end{equation}
Integration of 
\begin{equation}
\sqrt{-g}=a^3\sin^2 \chi \sin\theta
\label{mesure}	
\end{equation}	
over $S^{3}$ yields its volume,
i.e.,
\begin{equation}
V=2\pi^2 a^3.
\label{volume}	
\end{equation}

Below, $0$, $1$, $2$, $3$ correspond to $t$, $\chi$, $\theta$, $\varphi$, respectively. 
The non vanishing components of the Levi-Civita connection are 
\begin{eqnarray}
&&	
\Gamma^{1}_{22} = - \mathrm{sin}\chi\,\mathrm{cos}\chi,
\hspace{0.5cm}\Gamma^{1}_{33}=-\mathrm{sin}\chi\,\mathrm{cos}\chi\, \mathrm{sin}^{2}\theta,
\hspace{0.5cm}\Gamma^{2}_{12}=\Gamma^{2}_{21}=\mathrm{cot} \chi,
\hspace{0.5cm}\Gamma^{2}_{33}=-\mathrm{cos} \theta \, \mathrm{sin} \theta,
\nonumber
\\
&&
\Gamma^{3}_{13}=\Gamma^{3}_{31}=\mathrm{cot} \chi,
\hspace{0.5cm}\Gamma^{3}_{23}=\Gamma^{3}_{32}=\mathrm{cot} \theta.
\label{connections}
\end{eqnarray}
It follows then the non vanishing components of the Ricci tensor
\[
R_{11} = -2,\hspace{0.5cm}R_{22} = -2\mathrm{sin}^2\chi,\hspace{0.5cm}R_{33} = -2 
\mathrm{sin}^2\chi\,\mathrm{sin}^2\theta
\]
leading to the Ricci curvature 
\begin{equation}
R=\frac{6}{a^2}.	
\label{curvature}	
\end{equation}
The associated non vanishing components of the Einstein tensor are
\[
G_{00} = -\frac{3}{a^2}, \hspace{0.5cm}
G_{11} = 1,\hspace{0.5cm}
G_{22} = \mathrm{sin}^2\chi,\hspace{0.5cm}
G_{33} =  
\mathrm{sin}^2\chi\, \mathrm{sin}^2\theta
\]

\subsection{Feynman propagator at finite $T$} 
\label{thermal-green}

The Feynman propagator satisfies

\begin{equation}
\left(\nabla_{\mu}\,\nabla^{\mu}+M^{2}+\xi R\right)G_{{\cal F}}({\rm x},{\rm x}')=-\frac{1}{\sqrt{-g}}\delta({\rm x}-{\rm x}')	
\label{wave3}
\end{equation}
with the operator acting at point x of the biscalar. 
In order to obtain $G_{{\cal F}}({\rm x},{\rm x}')$ at finite temperature $T$, we take $\tau:=it$ in eq. (\ref{le}) and consider it real  with period $\beta=1/T$ in eq. (\ref{wave3}) \cite{dav82}. Then, noticing eqs. (\ref{mesure}), (\ref{connections}) and (\ref{curvature}), eq. (\ref{wave3}) yields
\begin{equation}
\left(	
-\frac{\partial^2}{\partial \tau^2}-\frac{1}{a^2}\Delta+M^2+\frac{6\xi}{a^2}
\right)G_{E}({\rm x},{\rm x}')=
\frac{\delta(\chi-\chi')\delta(\theta-\theta')\delta(\varphi-\varphi')}{a^3\sin^2 \chi \sin\theta}
\sum_{m=-\infty}^{\infty}\delta(\tau-\tau'-m\beta)
\label{propagator}
\end{equation}
where
\begin{equation}
	G_{E}({\rm x},{\rm x}'):=iG_{{\cal F}}({\rm x},{\rm x}')
	\label{feynman}
\end{equation}
and
\begin{equation}
\Delta:=\frac{1}{\sin^2\chi}\left[\frac{\partial}{\partial\chi} \left(\sin^2\chi\frac{\partial}{\partial\chi}\right) +\frac{1}{\sin\theta}\frac{\partial}{\partial\theta}\left(\sin\theta
\frac{\partial}{\partial\theta}\right)+
\frac{1}{\sin^2\theta}\frac{\partial^2}{\partial\varphi^2}
\right]	
\label{laplacian}	
\end{equation}
is the Laplacian on $S^3$ whose eigenfunctions are the 
spherical harmonics $W_{p\, n}^{k}(\chi,\theta,\phi)$
on $S^3$. That is
\begin{equation}
-\Delta W_{p\, n}^{k}(\chi,\theta,\phi)=p(p+2) 
W_{p\, n}^{k}(\chi,\theta,\phi)
\label{efunction}	
\end{equation}
where
\begin{equation}
W_{p\, n}^{k}(\chi,\theta,\phi)= 
\left[\frac{2^{2n+1}(p+1)\Gamma(p-n+1)\Gamma^2(n+1)}{\pi\Gamma(p+n+2)}
\right]^{1/2}\sin^n\chi C_{p-n}^{(n+1)}(\cos\chi)
Y_{n}^{k}(\theta,\phi)
\label{harmonics}	
\end{equation}
$p=0,1,\cdots\geq n=0,1,\cdots$; $k\in \{-n,-n+1,\cdots,n-1,n\}$;
$Y_{n}^{k}(\theta,\phi)$ are the well known spherical harmonics on $S^2$;
and $C_{l}^{(\omega)}({\rm x})$ is solution of the ultraspherical equation
(see eq. (13.81) in ref. \cite{arf85}).
The factors between brackets on the r.h.s. of eq. (\ref{harmonics}) are such that the eigenfunctions are orthonormal, namely
\begin{equation}
\int_{0}^{\pi}d\chi\int_{0}^{\pi}d\theta\int_{0}^{2\pi}d\varphi\;
W_{p\, n}^{k}(\chi,\theta,\phi)\,	
W_{p'\, n'}^{*\,  k'}(\chi,\theta,\phi)\sin^2 \chi \sin\theta
=\delta_{p\, p'}\delta_{n\, n'}\delta_{k\, k'}
\label{orthonormality}	
\end{equation}
It turns out that the set of eigenfunctions $W_{p\, n}^{k}$ is a basis on $S^3$. 
Indeed, the following closure relation is satisfied
\begin{equation}
\sum_{p=0}^{\infty}\sum_{n=0}^{p}\sum_{k=-n}^{n}
W_{p\, n}^{k}(\chi,\theta,\phi)\,	
W_{p\, n}^{*\,  k}(\chi',\theta',\phi')
=
\frac{\delta(\chi-\chi')\delta(\theta-\theta')\delta(\varphi-\varphi')}
{\sin^2 \chi \sin\theta}
\label{closure}	
\end{equation}

By defining
\begin{equation}
\psi_{m,p,n,k}({\rm x}):=\frac{1}{\sqrt{\beta a^3}}
W_{p\, n}^{k}(\chi,\theta,\phi)e^{i2\pi m\tau/\beta},\hspace{0.4cm}
E_{m,p}:=\frac{(p+1)^2}{a^2}+M^2+\frac{6\xi-1}{a^2}+\frac{4\pi^2 m^2}{\beta^2}
\label{eigenvalue}
\end{equation}
and using eq. (\ref{efunction}),
we can easily check that $\psi_{m,p,n,k}$ is eigenfunction of
the operator between brackets in eq. (\ref{propagator}) with eigenvalue $E_{m,p}$. Now we propose Schwinger's proper time representation of the finite temperature Green function \cite{deu79}
\begin{equation}
G_{E}({\rm x},{\rm x}')=
i\hspace{-0.2cm}\sum_{m=-\infty}^{\infty}
\sum_{p=0}^{\infty}\sum_{n=0}^{p}\sum_{k=-n}^{n}
\int_{0}^{\infty}d\omega
e^{-i\omega E_{m,p}}\psi_{ m,p,n,k}({\rm x})
\psi^{*}_{m,p,n,k}({\rm x}')
\label{ge}
\end{equation}
with $M^2$ in eq. (\ref{eigenvalue}) carrying  an infinitesimal negative imaginary part to make the integration over $\omega$ in eq. (\ref{ge}) to converge \cite{dew65}.

Indeed it follows that
\begin{eqnarray}
	&&
	\left(	
	-\frac{\partial^2}{\partial \tau^2}-\frac{1}{a^2}\Delta+M^2+\frac{6\xi}{a^2}
	\right)G_{E}({\rm x},{\rm x}')=-
	\sum_{m=-\infty}^{\infty}
	\sum_{p=0}^{\infty}\sum_{n=0}^{p}\sum_{k=-n}^{n}
	\psi_{ m,p,n,k}({\rm x})
	\psi^{*}_{m,p,n,k}({\rm x}')
	\nonumber
	\\
	&&
	\hspace{4.0cm}\times\int_{0}^{\infty}d\omega\frac{d}{d\omega}
	e^{-i\omega E_{m,p}}
	\label{propertime}
\end{eqnarray}
where the integration over $\omega$ is just minus unity. Therefore,
by using  Poisson's formula
\begin{equation}
	\frac{1}{2\pi}\sum_{l=-\infty}^{\infty} e^{-il\lambda}=\sum_{l=-\infty}^{\infty}\delta(\lambda-2\pi l)
	\label{poisson}
\end{equation}
and eq. (\ref{closure}),
it results that the r.h.s. of eq. (\ref{propertime}) is 
that in eq. (\ref{propagator}) as required.

We can obtain a more workable expression for the thermal Green function in eq. (\ref{ge}). Summations over $k$ and $n$  are evaluated by using the ``summation theorems'' in eq. (12.197) of ref. \cite{arf85},
and in eq. (8.934) of ref. \cite{gra07}, respectively. Then, considering further eqs. (8.936) and (8.937) in ref. \cite{gra07}, it results
\begin{eqnarray}
	G_{E}({\rm x},{\rm x}')=
	\frac{iT}{2\pi^2a^3\sin\alpha}\int_{0}^{\infty}d\omega\,
	e^{-i\omega \mu^2}
	\sum_{m=-\infty}^{\infty}e^{-i\omega(2\pi mT)^2+i2\pi mT(\tau-\tau')}
	\sum_{l=1}^{\infty}e^{-i\omega l^2/a^2}l\sin\, l\alpha 
	\nonumber
	&&
\end{eqnarray}
with $\alpha$ given in eq. (\ref{angle}),
\begin{equation}
\mu:=\sqrt{M^2+\frac{6\xi-1}{a^2}}
\label{mu}
\end{equation}
and the summation over $p$ replaced by that over $l:=p+1$, which after further manipulation clearly yields
\begin{eqnarray}
	G_{E}({\rm x},{\rm x}')=
	\frac{T}{4\pi^2a^3\sin\alpha}\int_{0}^{\infty}d\omega\,
	e^{-i\omega \mu^2}
	\sum_{m=-\infty}^{\infty}e^{-i\omega(2\pi mT)^2+i2\pi mT(\tau-\tau')}
	\sum_{n=-\infty}^{\infty}n e^{(-i\omega n^2/a^2)+i\alpha n}
	\nonumber
	&&
	\\
	\label{ge2}
	&&
\end{eqnarray}
At this point, by using eq. (\ref{poisson}), we have the following 
identities
\begin{eqnarray}
	&&
	\sum_{m=-\infty}^{\infty}e^{-i\omega(2\pi mT)^2+i2\pi mT(\tau-\tau')}=
	\frac{1}{2\pi}
	\sum_{m=-\infty}^{\infty}\int_{-\infty}^{\infty}d\lambda\,
	e^{-i\omega T^2\lambda^2+i\lambda T(\tau-\tau'-m\beta)}
	\nonumber
	\\
	&&
	\sum_{n=-\infty}^{\infty}n e^{(-i\omega n^2/a^2)+i\alpha n}=
	\frac{1}{4\pi^2}
	\sum_{n=-\infty}^{\infty}\int_{-\infty}^{\infty}d\lambda\,\lambda\,
	e^{-i\omega (\lambda/2\pi a)^2+i\lambda (\alpha-2\pi n)/2\pi}
	\nonumber
\end{eqnarray}
where the integrations over $\lambda$
can be promptly evaluated (see, e.g., ref. \cite{gra07})
to recast eq. (\ref{ge2}) as 
\begin{eqnarray}
	G_{E}({\rm x},{\rm x}')=-\frac{i}{\sin\alpha}
	\sum_{m=-\infty}^{\infty}\sum_{n=-\infty}^{\infty}
	(\alpha-2\pi n)\int_{0}^{\infty}\frac{d\omega}{(4\pi\omega)^2}
	e^{i\{[(\tau-\tau'-m\beta)^2+(\alpha-2\pi n)^2 a^2]/4\omega-\mu^2\omega\}}
	\nonumber
	&&
	\\
	\nonumber
	&&
\end{eqnarray}
The last step is to solve the integration over $\omega$ above (see, e.g., refs. \cite{gra07} or \cite{mat17}) to obtain an expression for $G_{E}$ [and for $G_{{\cal F}}$ by means of eq. (\ref{feynman})] carrying modified Bessel functions of the second kind
\begin{eqnarray}
	G_{E}({\rm x},{\rm x}')=\frac{\mu}{4\pi^2\sin\alpha}
	\sum_{m=-\infty}^{\infty}\sum_{n=-\infty}^{\infty}(\alpha-2\pi n)
	\frac{K_{1}\left(\mu\sqrt{(\tau-\tau'-m\beta)^2+(\alpha-2\pi n)^2 a^2}\right)}{\sqrt{(\tau-\tau'-m\beta)^2+(\alpha-2\pi n)^2 a^2}}.
	\label{ge3}
\end{eqnarray}

\vspace{1cm}
\noindent{\bf Acknowledgements} -- 
The
work of E. S. M.  has been partially supported by
``Funda\c{c}\~{a}o de Amparo \`{a} Pesquisa do Estado de Minas Gerais'' (FAPEMIG)
and by ``Coordena\c{c}\~{a}o de Aperfei\c{c}oamento de Pessoal de N\'{\i}vel Superior'' (CAPES). J. P. A. P. acknowledges a grant from
``Coordena\c{c}\~{a}o de Aperfei\c{c}oamento de Pessoal de N\'{\i}vel Superior'' (CAPES).


\begin{thebibliography}{88}
	
\bibitem{pen65}	
A. A. Penzias and R. W. Wilson,
A Measurement of Excess Antenna Temperature at 4080 Mc/s,	
Astrophys. J. {\bf 142}, 419 (1965).


	
\bibitem{par69}	
L. Parker, 
Quantized Fields and Particle Creation in Expanding Universes. I,
Phys. Rev. {\bf 183}, 1057 (1969).	



\bibitem{for75}
L. H. Ford, 
Quantum vacuum energy in general relativity, 
Phys. Rev. {\bf 11}, 3370 (1975).
	



\bibitem{sch38}
E. Schr\"{o}dinger, 
{\it Eigenschwingungen des sph\"{a}rischen Raumes},
Comment. Pont. Acad. Sci. {\bf 2}, 321 (1938).


\bibitem{str75}
E. Streeruwitz, 
Vacuum fluctuations of a scalar field in an Einstein universe, 
Phys. Lett. B {\bf 55}, 93 (1975).

\bibitem{mam76}
S. G. Mamaev, V. M. Mostepanenko and A. A. Starobinsky,
Particle creation from the vacuum near a homogeneous
isotropic singularity,
Sov. Phys. JETP {\bf 43}, 823 (1976).



\bibitem{dav82} 
N. D. Birrel and P. C. W. Davies,
\emph{Quantum Fields in Curved Space},
Cambridge University Press, Cambridge UK (1982).
	
	
\bibitem{ozc06}	
M. \"{O}zcan,
Casimir energy density for spherical universes in
{$n$}-dimensional spacetime,
Class. Quantum Grav. {\bf 23}, 5531 (2006). 
	
	
	
\bibitem{dow77}
J. S. Dowker and R. Critchley, 
Vacuum stress tensor in an Einstein universe: Finite-temperature effects, 
Phys. Rev D {\bf 15}, 1484 (1977).
	
	
	
\bibitem{alt78}
M. B. Altaie and J. S. Dowker, 
Spinor fields in an Einstein universe: Finite-temperature effects, 
Phys. Rev D {\bf 18}, 3557
(1978).
	
	
\bibitem{bro69}
L. S. Brown and G. J. Maclay, 
Vacuum Stress between Conducting Plates: An Image Solution, 
Phys. Rev. {\bf 184}, 1272 (1969).

\bibitem{kap06}
J. I. Kapusta and C. Gale, {\it Finite-Temperature Field Theory Principles and Applications}, 
Cambridge University Press, Cambridge UK (2006).

\bibitem{hua87}
K. Huang, 
\emph{Statistical Mechanics}, 
John Wiley \& Sons, U.S.A. (1987).

\bibitem{bre02}
I. Brevik, K. A. Milton and S. D. Odintsov,
Entropy Bounds in $R \times S^3$ Geometries,
Ann. Phys. (Amsterdam) {\bf 302}, 120 (2002).		

\bibitem{eli03}
E. Elizalde and A. C. Tort, 
Entropy bounds for massive scalar field in positive curvature space,
Phys. Rev. D {\bf 67}, 124014 (2003).		


\bibitem{bez11}
V. B. Bezerra, G. L. Klimchitskaya, V. M. Mostepanenko and C. Romero,
Thermal Casimir effect in closed Friedmann universe revisited,
Phys. Rev. D {\bf 83}, 104042 (2011).		


\bibitem{her06}
C. A. R. Herdeiro and M. Sampaio, Casimir energy and a cosmological bounce, 
Class. Quantum Grav. {\bf 23}, 473 (2006).


\bibitem{cal85} 
H. B. Callen, 
\emph{Thermodynamics and an Introduction to Thermostatistics},
John Wiley \& Sons, U.S.A. (1985).

\bibitem{lor15}
V. A. De Lorenci, L. G. Gomes and E. S. Moreira Jr., Hot scalar   
radiation setting bounds on the curvature coupling parameter, 
Class. Quantum Grav. {\bf 32},   085002   (2015).


\bibitem{mor17}
E. S. Moreira Jr., Hot scalar radiation around a cosmic string   
setting bounds on the coupling parameter $\xi$, 
JHEP {\bf 03}, 105 (2017).

\bibitem{mor20}
E. S. Moreira Jr., Ambiguities in the local thermal behavior of   
the scalar radiation in one-dimensional boxes, 
Phys. Rev. D {\bf 102}, 085014 (2020).


\bibitem{arf85} 
G. Arfken, 
\emph{Mathematical Methods for Physicists}, 
Academic Press, USA  (1985).


\bibitem{abr65} M. Abramowitz  and I. Stegun, 
\emph{Handbook of Mathematical Functions with Formulas, Graphs, and Mathematical Tables}, 
Dover Publications, USA  (1965).


\bibitem{ful89} 
S. A. Fulling,  
\emph{Aspects of Quantum 
	Field Theory in Curved Space-Time}, 
Cambridge University Press, Cambridge UK (1989).




\bibitem{gra07} I. S. Gradshteyn and I. M. Ryzhik, 
\emph{Table of Integrals, Series, and Products}, 
Academic Press, USA  (2007).

\bibitem{mat17}
Wolfram Research, Inc.,
\emph{Mathematica, Version 14.1}, Champaign, IL (2024).





\bibitem{fer24} 	
E. J. B. Ferreira and H. F. Santana Mota,
Quantum Brownian motion induced by a scalar field
in Einstein’s universe,	
Eur. Phys. J. C. {\bf 84}, 412 (2024).	

\bibitem{med01}
R. Medina and E. S. Moreira Jr.,
Thermal fluctuations of a quantized massive scalar field in a Rindler background,
Phys. Rev. D {\bf 63}, 124022 (2001).






\bibitem{zhu96}
A. Zhuk and H. Kleinert, Casimir effect at nonzero temperatures in a closed Friedmann universe, Theor.  Math. Phys. {\bf 109}, 1483 (1996).

	
\bibitem{alt03}
M. B. Altaie and M. R. Setare, 
Finite-temperature scalar fields and the cosmological constant in an Einstein universe, 
Phys. Rev. D {\bf 67}, 044018 (2003).	
	
\bibitem{hu82}
B. L. Hu,
Finite temperature quantum fields in expanding universes.
Phys. Lett. B {\bf 108}, 19 (1982).



\bibitem{wel81}
H. E. Haber and  H. A. Weldon,
Thermodynamics of an Ultrarelativistic Ideal Bose Gas,
Phys. Rev. Lett. {\bf 46}, 1497 (1981).

	
\bibitem{deu79}
D. Deutsch and P. Candelas,
Boundary Effects in Quantum Field Theory,
Phys. Rev. D {\bf 20}, 3063 (1979). 


\bibitem{dew65}
B. S. DeWitt,
\emph{Dynamical Theory of Groups and Fields},
Blackie \& Son, London (1965).
	

	

	
\end{thebibliography}
\end{document}